\documentclass[12pt,twocolumn,journal]{IEEEtran}

\usepackage{subfigure}
\usepackage{graphicx}
\usepackage{amsmath}
\usepackage{tabularx}
\usepackage{array,ragged2e}
\usepackage{graphics}
\usepackage[dvips]{epsfig}
\usepackage[hyphens]{url}
\usepackage{hyperref}
\hypersetup{breaklinks=true}

\begin{document}

\title{Guidelines Towards Information-driven Mobility Management}
\author{

  Rute C. Sofia$^{1,2}$

\IEEEauthorblockA {$^{1}$ COPELABS - University Lusofona, Campo Grande 388, 1700-097 Lisboa, Portugal}

\IEEEauthorblockA{$^{2}$ ISTAR-IUL, ISCTE.}

}

\maketitle

\begin{abstract}
The architectural semantics of \emph{Information-Centric Networking} bring in interesting features in regards to mobility management: Information-Centric Networking is content-oriented, connection-less, and receiver-driven. Despite such intrinsic advantages, the support for node movement is being based on the principles of IP solutions. IP-based solutions are, however, host-oriented, and Information-Centric Networking paradigms are information-oriented.
By following IP mobility management principles, some of the natural mobility support advantages of Information-Centric Networking are not being adequately explored. 
 This paper contributes with an overview on how Information-Centric Networking paradigms handle mobility management as of today, highlighting
current challenges and proposing a set of design guidelines to overcome them, thus steering a vision towards a content-centric mobility management approach.
\end{abstract}





\section{Introduction}
Internet traffic is consumed and produced by heterogeneous
sets of mobile, resource-constrained end-user devices which are interconnected via a fixed or wireless/cellular infrastructures. Moreover, the evolution of the
Internet infrastructure, of which 5G is a relevant part, brings in
new requirements, such as the need to support large-scale \emph{Internet of Things (IoT)} environments; strict end-to-end latency requirements;  service-oriented model support\cite{Ravindran}. 
 Mobility management plays a key part in this evolutionary step of the Internet, and IP-based mobility solutions have been evolving towards the support of network decentralisation, to be able to cope with high topological variability, among other issues. Being based on the principles of IP networks only, current mobility management solutions face limitations such as, for instance, the lack of integrated security; the need for an end-to-end path between consumers and producers; being focused on host reachability, instead of on data reachability.
Several engineering workarounds have been assisting the evolution of such mobility management solutions towards more complex, large-scale environments. 

\emph{Information-Centric Networking (ICN)} architectures such as the \emph{Named Data Networking (NDN)} architecture, have intrinsic features that are better suited to support environments with a high degree of mobility. For instance, ICN focuses on content and not on hosts as the addressable
entities, thus providing better communication support while devices are on the move. Its connection-less nature and interface abstraction model are interesting features to support many-to-many communications, even if connectivity is intermittent~\cite{Afanasyev2016}. Its per packet pull-based communication model is, at a first sight, sufficient to support consumer node mobility. On the other hand, its pull-based receiver-driven model does not support well mobility of producer nodes, as shall be explained further in section 4.4. Producer mobility is being handled by anchor-based proposals that mimic, in some aspects, IP-based mobility management and consequently, are following a host-reachability mobility management model, instead of a content-centric one. 

To better understand how to develop future mobility management solutions, it is necessary to think about the different mobility management functions, and how they are served (or not served) by ICN.

This work contributes to the debate on how to evolve mobility management, in a way that truly becomes content-centric:
\begin{itemize}
    \item To highlight the functions that compose mobility management, based on the main architectural solutions developed so far (section 3).
    \item To explain ICN mobility management efforts, highlighting challenges to overcome (sections 4, 5).
    \item To provide a set of architectural guidelines aiming at providing a content-centric approach to mobility management and yet, assisting interoperability needs (sections 5, 6).
\end{itemize}

For this purpose, section \ref{sec:relatedwork} covers related work explaining our contributions, while section \ref{sec:mobility management} provides a debate on mobility management functional aspects. Section \ref{sec:mobilityNDN}
covers ICN mobility management. Guidelines towards a content-centric
mobility management solution are provided in section \ref{sec:guidelines},
being the paper concluded in section~ \ref{sec:conclusions}, where future directions for research on this topic are also provided.

\section{Related Work}

\label{sec:relatedwork}

Mobility management comprises a wide set of related work, including an extensive set of proposals that has been developed to support mobility from the perspective of different OSI layers~\cite{Chen2016mobility}.  Out of the available solutions, IP-based solutions are today the basis of mobility management in cellular and
wireless environments. The most recent evolution of such category of solutions concerns distributed mobility management and
is being steered by the \emph{Internet Engineering Task Force (IETF)} \emph{Distributed
Mobility Management (DMM)} Working Group~\cite{dmm}. 
Decentralisation of IP-based mobility management relates mostly with the integration of these approaches in large-scale heterogeneous environments (such as 5G) as well as with support towards flatter networking architectures~\cite{Liu2014}. 
The debate on decentralisation covers a wide set of topics, including decoupling of data and control planes; better management of mobility anchor points, etc.

ICN introduced a relevant simplification, namely: information-centricity instead of host-reachability. The capability to store status and data in routers (\emph{store-and-forward}) provides the grounds to better support mobility of devices in a network. In this context, a thorough overview on mobility
aspects for one of the existing ICN architectures,\emph{Named Data Networking (NDN)}, has been provided by Zhang et al.~ \cite{Afanasyev2016}. The authors approach advantages and disadvantages in different scenarios with the aim of further assisting the support of mobility. Their analysis is compared to IP-based approaches in terms of architectural design. Zhu et al. provide a global overview on the NDN design and mobility
support, alerting to the need to consider a better support for producer mobility \cite{Zhu2013a}. In fact, most related work has been focused on improving producer mobility, i.e., supporting movement of devices that provide data. Auge et al. provide a relevant overview on mobility support in particular for environments focused on the interoperability of ICN and IP, proposing an anchor-less solution to support mobility coupled to a routing protocol ~\cite{Auge2018}. Kite is a mobility solution for NDN which exploits NDN forwarding state to keep track of moving producers and their whereabouts. Kite follows IP-based approaches by considering a ``Rendez-Vous point'' which assists in tracing where data is, while the producer performs reattachment to a new location~\cite{Zhang2014}. 

Chen et al. describe steps towards a reference model for mobility-driven networks, debating on evolutionary principles such as the decoupling of service and device entity, for vertical handovers, and entity/locator identifiers, for horizontal handovers~\cite{Chen2014}.
Tyson et al. provide a survey on ICN mobility issues from an architectural perspective, highlightig potential benefits brought by the ICN networking semantics\cite{tyson2012survey}. 
Our paper closely follows the line of work that is focused on assisting in further evolving mobility management in an interoperable way, by learning from prior approaches, while at the same time by trying to keep the beneficial properties of ICN design (content-centricity). A contribution of our work is a clarification on different functional aspects of an abstract model of mobility management derived from prior learning. A second contribution is a clarification on the different functional entities (mobile node, correspondent node) and where they fit ICN architectures. A third contribution concerns an analysis of current mobility management approaches, and guidelines to assist a consolidated design of future mobility management approaches.
\section{Mobility Management Functional Aspects}

\label{sec:mobility management}

Mobility management is a relevant network function in today's Internet, and yet it is still one of the most challenging. The purpose
of mobility management is to provide support for active communication
in a way that allows services to be active with the least interruption,
while users are on the move. 
For that purpose, mobility management handles three main
processes: i) location management; ii) handover management; iii) multi-homing.

\emph{Location management }has as main purpose to allow data to flow
adequately between source and destinations, independently of the whereabouts
of the involved devices. Location management is supported by binding
mechanisms, that support the mapping between mobile nodes to specific
identifiers, both before, during, and after a move occurs.

\emph{Handover management} concerns being able to identify new points
of attachment for mobile nodes, and to allow data and signalling to
flow to the new whereabouts of devices, while these are moving. 

From an end-user perspective, \emph{Multi-homing} concerns support for a device to use simultaneously its multiple interfaces, in order to increase performance and/or reliability of data transmission. From a network perspective, multi-homing concerns supporting one or multiple services, via two or more distinct network regions (or segments), towards consumers.

\begin{figure}
\centering
\includegraphics[scale=0.35]{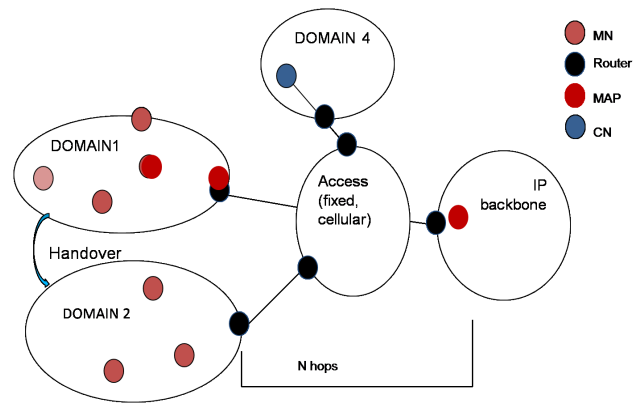}

\caption{High-level representation of mobility management main entities, the MN, the CN, and the MAP. The
MAP is often co-located with different devices, including end-user
devices.\label{fig:Mobility-management-functional}}

\end{figure}

In a pursuit to support these three processes, IP-based mobility management solutions share three main functional entities: i) \emph{Mobile Nodes}; ii) \emph{Correspondent Nodes}; iii) \emph{Mobility Anchor Points}. The placement of this functional entities is illustrated in Figure
\ref{fig:Mobility-management-functional}. Such entities can then
be co-located with different devices, depending on the selected mobility management approach \cite{Nascimento2011}. 

The \emph{Mobile Node 
(MN)} corresponds to a functional entity that is part of an end-user
device. Today, it is often located in a portable, battery-constrained
device which is wireless or cellular enabled. The MN is the mobile or static entity that triggers communication.

The MN has an active
communication towards peers over the internet, known as the MN \emph{Correspondent
Nodes (CNs)}. The MN has one (or more)
identifiers, i.e., IPv6 addresses such as occurs in \emph{Mobile IPv6
(MIPv6)} and its extensions; URIs for a mobility management solution
such as the \emph{Session Initiation Protocol (SIP}); a locator-id
based identifier for a solution such as the \emph{Host Initiation
Protocol (HIP)}. The MN functional role resides both on the data and
control plane.

The CN represents an \textquotedblleft active partner\textquotedblright{}
of the MN. The CN as defined in IETF RFC4885 \cite{rfc4885} is "Any node that is communicating with one or more MNs.  A CN could be either located within a fixed network or within a mobile network, and could be either fixed or mobile.". Today, it is an entity residing in a mobile device and is the receiver of a communication process started by the MN. In ICN, the functional representation of MN vs. CN could be simplified by considering a single entity, as we shall further explain in sections 4 and 5.
The reason to still differentiate between these two functional entities relates with the evolution of the Internet: at first, mobility management approaches were developed to support service and session continuity for the consumer of that service. This was the MN. The signalling required to support handovers was devised having in mind that particular entity, and assuming that all other elements in the network would be static. Later, with the introduction of two-way real-time communication in mobile environments, the solutions developed integrated extensions to handle CN mobility, as well as to attempt to handle simultaneous mobility by all of the involved parties.

The \emph{Mobility Anchor point (MAP)} is a functional control plane
entity that may reside in a network element (e.g., in a router) or
in an end-user element (e.g., end-user equipment, server). The MAP
controls the main functionality of mobility management, namely:
handover management; traffic offloading; location management; bindings
and address translation; \emph{Quality of Service (QoS)} and forwarding
policies.

Learning from the evolution of prior solutions, and from the extension required to support additional features such as simultaneous mobility, it is relevant to consider that any future mobility management architecture needs to be designed already having in mind that any node on the network can move. Furthermore, it needs to consider that due to the way the Internet is evolving, these functional entities can reside in any type of device, even embedded ones.

To further debate on how such support can be provided, the functional design
of today's solutions can be split into different blocks \cite{Nascimento2011}:
\begin{itemize}
\item \emph{Identification. }For IP-based solutions such as \emph{MIPv6},
this would correspond to the mapping of a network interface to an
IPv6 address; in SIP this would be a mapping between an URI (known
address) and one or multiple IPs; in HIP this corresponds to a Locator
Id.
\item \emph{Database control}, control functionality usually provided by
a central entity, which assists in a quicker mapping of the different
identifiers. Usually, this functionality is part of the MAP entity.
\item \emph{MAP selection}, which is a control function that assists in
a better deployment of MAPs having in mind to improve reachability
of MNs. In centralised solutions, such selection is often performed
in a static and centralised way.
\item \emph{Binding registration}, control plane function that signals the
first registration of a MN in a mobile system. For instance, in MIP
it is the first Binding Update message sent to a MAP or to a CN. In
SIP it is the REGISTER message sent to the Registrar server.
\item \emph{Binding update}, control plane function that signals an update
a record in the Identification control. Binding updates are used when
the unique identifier of a device changes. 
\item \emph{Routing or forwarding}: it is the process of intercepting the
packets destined to the known-address, encapsulating them with the
real-address, and forwarding them. In MIP this is performed by the
MAP or by the first access router in the path \emph{(Home Agent, HA);}
in SIP this process can be performed by an external element, for instance,
an RTP translator.
\item \emph{Handover negotiation}: the process taken when the device has
its identifier changed, to allow active communications to be held
with the least disruption. In MIP, the handover negotiation may be
anticipated with e.g., mechanisms such as the Fast Handover extension.
SIP does not implement any anticipation, performing a re-negotiation
after the connection between peers is lost. 
\item \emph{Resource management}, the process that assists in guaranteeing
the quality of a connection while devices perform a handover. Most
solutions as of today do not integrate QoS support, recurring to external
mechanisms to provide such support. 
\item \emph{Mobility }anticipation: it is the procedure of performing a
handover before an active connection experiences a break. For instance,
anticipation is partially supported in MIPv6 via the extension Fast
Handovers for MIP. 
\item \emph{Security and privacy:} it refers to every security mechanism
to assure the integrity of both data and channel for the active communication,
as well as for the signalling of the mobility management system. Current
centralised solutions require external security support to protect
data, channel, as well as involved signalling.
\end{itemize}
The evolution of mobility management towards information-centricity
(and hence, better service support) requires looking into these different
functional blocks, and understanding whether or not they can be simplified.
To further assist such evolution it is also relevant to remind that
IP-based mobility management approaches have been designed having
in mind support of mobility from a source-driven perspective. On later
phases, adjustments of the centralised solutions for support of simultaneous
mobility \cite{Wong2007} as well as non-simultaneous mobility have
been introduced. This is the case, for instance, of the \emph{Return
Routability Procedure} for MIPv6 solutions, intended to assist CN
mobility. 

As also stated in section 2., efforts towards the evolution of IP-based
mobility management is approaching a distributed vision, having in
mind the support of mobility for the different entities, where IPv6
is the underlying protocol. In such context, solutions have looked
into MAP selection and discovery; forwarding path and signalling management;
exchange of control information to assist faster handovers (e.g.,
better selection of identifiers to use on the new attachment locations).
\section{Mobility in ICN}
\label{sec:mobilityNDN}

\subsection{Mobile Nodes, Correspondent Nodes, and MAPs in ICN}
The ICN architecture embodies a publish/subscribe pull-based communication
model. Producer nodes correspond to devices that send data (\emph{Data
packets}), once they get an expression of interest by consumer nodes (\emph{Interest
packets}). 
Data is sent back following ICN forwarding strategies, and based on the network state left by Interest packets in routers along the way.

From a functional and interoperable mobility management perspective, 
producer nodes and consumer nodes may be associated with both the MN or the CN functionality.
At a first glance, and from an abstract, functional perspective, mobility management entities could be reduced from MN vs. CN into a single MN entity, for instance. However, ICN is receiver-driven, while IP mobility management solutions are source-driven. 
Furthermore, ICN does not require the functional concept of a MAP to support mobility, as binding is directly performed to content and not to hosts, as shall be explained next.

\subsection{Architectural Design Advantages}

ICN integrates several features that are beneficial from a mobility
management perspective. To assist in the understanding of such advantages,
Table \ref{tab:mobilityICN} provides an overview on how the different
mobility management functional blocks described in section \ref{sec:mobility management}
are supported via the most emblematic mobility management solutions
of today. 

From a mobility management perspective, a first advantage of the architectural design of ICN against its IP-based counterparts is the
focus on content, instead of on host reachability. In ICN paradigms content
becomes the addressable entity, instead of a host identifier. Content
is also the routing target, which serves better the handover process:
there is no need for a database identifier control process, for instance. 

A second advantage of ICN is its interface abstraction, \emph{Face}.
Faces provide a better support for multi-homing, including security
\cite{Schneider2015}. Faces are also relevant in the support of
distributed mobility management. The Face abstraction provides the
means for applications to seamlessly and securely interact with multiple
physical and virtual interfaces, as there is no dependency on interface
identifiers nor or on host identifiers. Adding to the Face abstraction,
Forwarding strategies serve better multi-homed devices, as Interest
packets can be forwarded having in mind specific requirements for
multi-homed environments. Forwarding strategies are based on the information
stored in the \emph{Forwarding Information Base (FIB)} and additional
traffic measurement. The \emph{Pending Interest Table (PIT)} stores
Name Prefixes for which consumers expressed interest. Data packets
simply follow the state left in the PITs.

\begin{table*}[]
\caption{Mobility management functional blocks, support in different mobility
management solutions .\label{tab:mobilityICN}}
\centering
\begin{tabular}{ 
|>{\RaggedRight\arraybackslash}p{2cm}|>{\RaggedRight\arraybackslash}p{2cm}|>{\RaggedRight\arraybackslash}p{2cm}|>{\RaggedRight\arraybackslash}p{2cm}|>{\RaggedRight\arraybackslash}p{2cm}|>{\RaggedRight\arraybackslash}p{2cm}| }
\hline
\textbf{\scriptsize{}Functional blocks} & \textbf{\scriptsize{}MIP~\cite{mipv6}} & \textbf{\scriptsize{}SIP~\cite{sip}} & \textbf{\scriptsize{}HIP~\cite{Nikander}} & \textbf{\scriptsize{}M-SCTP~\cite{koh2005}} & \textbf{\scriptsize{}ICN~\cite{Afanasyev2016}}\\
\hline
{\scriptsize{}Identification} & {\scriptsize{}IP address (interface)} & {\scriptsize{}URI (unique, associated with user)} & {\scriptsize{}Locator Id (device)} & {\scriptsize{}IP and port} & {\scriptsize{}Name prefix (content)}\tabularnewline
\hline 
{\scriptsize{}Id database control} & {\scriptsize{}MAP} & {\scriptsize{}Centralised, controlled by the provider. access through
the MAP} & {\scriptsize{}MAP} & {\scriptsize{}None} & {\scriptsize{}Not required}\tabularnewline
\hline 
{\scriptsize{}MAP} & {\scriptsize{}Centralised solution, located in the provider premises
(HA, access router) } & {\scriptsize{}Centralised solution, located in the provider premises
Proxy SIP (server) } & {\scriptsize{}Centralised solution, located in the provider premises } & {\scriptsize{}Centralised solution, located in the provider premises } & {\scriptsize{}Not required}\tabularnewline
\hline 
{\scriptsize{}Binding mechanism} & {\scriptsize{}Periodic Binding Update message, MN to HA, MAP or CN } & {\scriptsize{}REGISTER message, MN to Registrar Server or Outbound
Proxy } & {\scriptsize{}-} & {\scriptsize{}-} & {\scriptsize{}Pull-based Interest packet approach; in-network caching}\tabularnewline
\hline 
{\scriptsize{}Routing/  forwarding} & {\scriptsize{}IP based (shortest-path)} & {\scriptsize{}Proxy or RTP translator } & {\scriptsize{}Dual, based on locator and on IP } & {\scriptsize{}IP-based} & {\scriptsize{}Data-based routing, forwarding strategies adapted to
mobility}\tabularnewline
\hline 
{\scriptsize{}Handover negotiation} & {\scriptsize{}Make- before- break, with FMIP access routers negotiation } & {\scriptsize{}Break- before- make, RE- INVITE message, MN to CN } & {\scriptsize{}Make before break } & {\scriptsize{}Break before make, requires setup of new TCP connection } & {\scriptsize{}No need for consumers; required for producers}\tabularnewline
\hline 
{\scriptsize{}Resource management} & {\scriptsize{}None} & {\scriptsize{}None} & {\scriptsize{}None} & {\scriptsize{}None} & {\scriptsize{}Forwarding strategies}\tabularnewline
\hline 
{\scriptsize{}Security/privacy} & {\scriptsize{}Not integrated} & {\scriptsize{}Not integrated} & {\scriptsize{}Yes, intrinsic to HIP} & {\scriptsize{}Not integrated} & {\scriptsize{}Yes}\tabularnewline
\hline 
{\scriptsize{}Handover Anticipation} & {\scriptsize{}Partial, e.g., FMIPv6} & {\scriptsize{}No} & {\scriptsize{}No} & {\scriptsize{}No} & {\scriptsize{}No}\tabularnewline
\hline
\end{tabular}
\end{table*}

Thirdly, the pull-based communication model of NDN, where data is
only sent if Interest packets are first transmitted, allows for a
binding signalling reduction during the handover process. 

A fourth advantage of the architectural design proposed in ICN concerns
the flexible forwarding strategies and the routing, which is data-oriented.
Such approach provides better support for multi-homing environments,
as well as for the support of frequent movement by devices. 

\subsection{Mobility Management in Different ICN Solutions}
\begin{table*}[]
\caption{ICN main approaches, mobility challenges.\label{tab:ICN-main-approaches,-1}}
\centering
\begin{tabular}{ |>{\RaggedRight\arraybackslash}p{2cm}|>{\RaggedRight\arraybackslash}p{2cm}|>{\RaggedRight\arraybackslash}p{2cm}|>{\RaggedRight\arraybackslash}p{2cm}|>{\RaggedRight\arraybackslash}p{2cm}| }
\hline
\textbf{\scriptsize{}Approach} & \textbf{\scriptsize{}Mobility management description} & \textbf{\scriptsize{}Consumer Mobility} & \textbf{\scriptsize{}Producer Mobility} & \textbf{\scriptsize{}Multi-homing}\tabularnewline
\hline 
\hline 
{\scriptsize{}DONA~\cite{koponen2007}} & {\scriptsize{}Anchor-based, early-binding approach, producers register
identifier to locator mapping that must be resolved before data can
be sent. Intends to be interoperable with DNS.} & {\scriptsize{}Supported, but not intrinsic} & {\scriptsize{}No} & {\scriptsize{}No}\tabularnewline
\hline 
{\scriptsize{}CCNx~\cite{Gusev2017}} & {\scriptsize{}Anchor-less, late binding approach, as data is only sent after an Interest packet is received. There is no direct identifier
\textendash{} locator mapping CCNx can handle 97\% of requests during
high mobility. } & {\scriptsize{}Intrinsic. When a consumer moves, Interest packets are
again sent. } & {\scriptsize{}No} & {\scriptsize{}Yes}\tabularnewline
\hline 
{\scriptsize{}NetInf~\cite{Dannewitz2013}} & {\scriptsize{}Anchor-based, early-binding, similar to DONA, even though
it requires consumer lookups} & {\scriptsize{}Supported but not intrinsic.} & {\scriptsize{}No} & {\scriptsize{}No}\tabularnewline
\hline 
{\scriptsize{}PSIRP~\cite{dimitrov2010}} & {\scriptsize{}Anchor-based, late-binding, requires consumer re-registration
after moving. } & {\scriptsize{}Intrinsic. When a consumer moves, Interest packets are
again sent. } & {\scriptsize{}No} & {\scriptsize{}No}\tabularnewline
\hline 
{\scriptsize{}JUNO~\cite{Tyson2012}} & {\scriptsize{}Middleware takes care of information-centric functionality.
Relies on a DHT approach, where flat identifiers for content are registered.} & {\scriptsize{}Supported but not intrinsic. Middleware takes care of
the mobility. } & {\scriptsize{}No} & {\scriptsize{}Yes}\tabularnewline
\hline 
{\scriptsize{}NDN~\cite{zhang2010}} & {\scriptsize{}Similar to CCNx.} & {\scriptsize{}Intrinsic. When a consumer moves, Interest packets are
again sent.} & {\scriptsize{}No} & {\scriptsize{}Yes}\tabularnewline
\hline 
\end{tabular}{\scriptsize \par}
\end{table*}

While ICN approaches have in common the advantages described in the
previous section, different approaches tackle mobility management in different ways. To assist in a better understanding of the current situation, Table \ref{tab:ICN-main-approaches,-1} describes whether/how producer mobility, consumer mobility, and multi-homing are supported
by reference approaches.

As described in Table 2, none of the main ICN architectures provides seamless producer mobility support. As for consumer mobility, existing approaches either take care of
such support based on anchor-based approaches, following IP-based
learning, or via anchor-less approaches. Furthermore, multi-homing is supported only in JUNO, CCNx and NDN. 

These aspects are further debated on the next sections. The description provided is focused on the line of work derived from CCNx, including NDN.

\subsection{Multi-homing }

ICN supports end-user device multi-homing via the Face abstraction. Support for networked multi-homing is achievable via ICN multi-path forwarding strategies, and routing. The Face abstraction brings in the possibility to jointly explore data transfer to multiple services and applications as well as to physical interfaces. Moreover, multi-homing is supported with fine-grained control: the ICN per packet pull-based model provides better support for resource management aspects, such as
load-balancing (based on packets instead of flows). 
The ICN forwarding strategies are applied on a local basis: different nodes and/or regions of nodes can have different forwarding strategies, which strengthens multi-homing support capability via ICN.

Therefore, multi-homing is a mobility management process that is naturally
supported by ICN approaches. In comparison, prior solutions required additional support of, for instance, \emph{Quality of Service (QoS)} mechanisms.

\subsection{Consumer Mobility}
In order to explain how ICN supports consumer mobility, this section provides two examples: MN acting as consumer; CN acting as consumer. The explanation provided is based on the functional entities described in section \ref{sec:mobility management}, which are the basis for today's mobility management reference architectures, onto ICN. The purpose is to explain limitations that may arise from such mapping.
$MN$ is an ICN producer that is directly connected to the NDN router $B$, and in active communication
with a consumer $CN$. Both MN and CN reside in mobile nodes. Connectivity can be intermittent.

\subsubsection{MN as Consumer}

\begin{figure}
\centering
\includegraphics[scale=0.5]{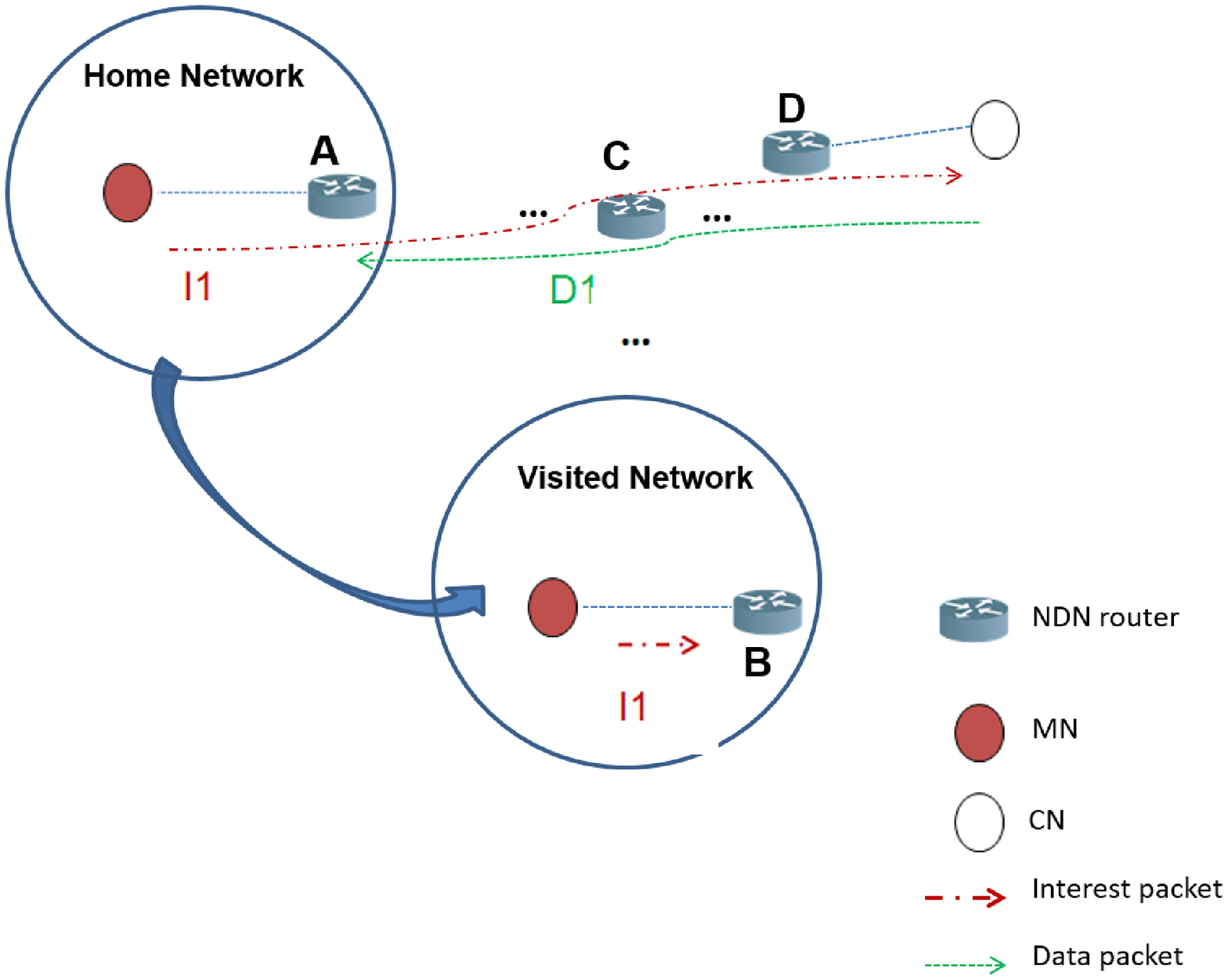}\caption{MN as consumer node. The MN starts by expressing interest in content (Interest packet I1). This packet reaches the CN, which replies with data packet D1. In the meanwhile the MN moves, and does not receive D1. Upon reattachment, the MN sends I1 again. \label{fig:Mobility-ICNMN}}
\end{figure}

Figure~\ref{fig:Mobility-ICNMN} illustrates a scenario where a MN is attached to its original network, the \emph{home network}. A, B, C and D represent routers. 
As ICN is receiver-oriented, data transmission for this example starts when the consumer entity expresses interest on a specific content, i.e., MN sends an Interest packet $I1$ with a specific Name Prefix. The Interest packet is stored in the PIT of ICN devices along the path (routers A, C, D),until it reaches a node that has the requested content, the producer, or a router that already cached the respective content in its \emph{Content Store (CS}).
In the meanwhile, and while packet I1 is being transmitted, MN starts to move and reattaches to router B, that serves a \emph{Visited Network}. Based upon ICN principles, once reattachment occurs, MN again sends an Interest packet with the same Name Prefix (I1). When this packet reaches router C, this router already has the content requested stored (D1) and therefore, the forwarding of Interest packet I1 stops. The subsequent data exchange is directly handled between MN and any device that holds content requested by MN.

\subsubsection{CN as Consumer}

\begin{figure}
\centering
\includegraphics[scale=0.5]{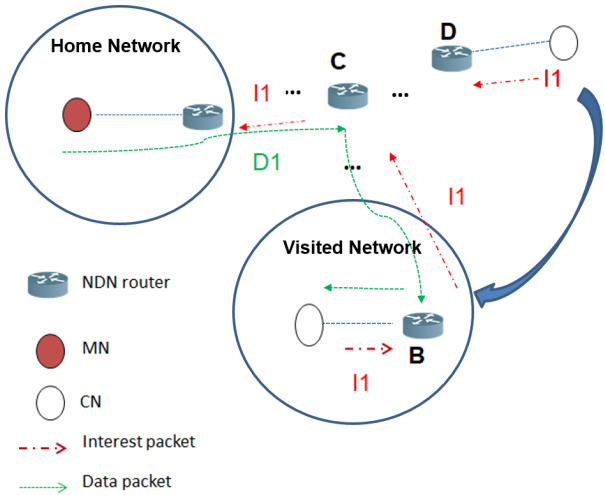}\caption{CN as Consumer. CN expresses interest by sending packet I1 and then moves. On the new location, CN again emits I1. Therefore, subsequent data packets reach CN at the visited network.\label{fig:Mobility-ICNCN}}
\end{figure}
 
 In this second example, illustrated in Figure \ref{fig:Mobility-ICNCN}, the CN entity is the consumer, while MN is a producer of information. In terms of consumer mobility, the situation is similar to the one described in the previous section: CN as a consumer expresses interest (sends an Interest packet, represented by I1) carrying a Name Prefix for specific content, which in our example is being produced by MN. In the meanwhile and either before receiving data, or already after receiving some data packets, CN moves to a new location, performing reattachment to NDN router B. The receiver-driven design of ICN implies that once CN reconnects to a new node (in our example, router B), it starts sending Interest packets to get the desired content, based on the respective application requirements and settings.
 Therefore, the pull-based receiver-driven nature of ICN is beneficial for the case of consumer mobility, independently of the entity that is moving. In other words: for the case of consumer mobility, there is no need to distinguish between a MN and a CN entity, in future mobility management solutions.
 
 \subsubsection{Consumer Mobility Discussion}
 
As the pull-based model relies on a per packet approach, even if a consumer already received some data chunks and then moves, transmission can be immediately re-established once the consumer (MN or CN)
can send data to a neighbour. In our examples, this is synonymous with the consumer being in a state that allows it to forward Interest packets again. 
At a first glance, ICN supports consumer mobility well. Nevertheless, in large-scale networks and environments where consumers
move frequently and fast (e.g., vehicular networks, personal Internet of Things environments) data transmission
may still be affected by frequent movement. Even though the in-network
caching provided by ICN can counter-balance such situations, the performance
of the data transmission is highly dependent on aspects such as the
type of topology, type of movement, and speed of nodes. 

\subsection{Producer Mobility}

\begin{figure}
\centering
\includegraphics[scale=0.5]{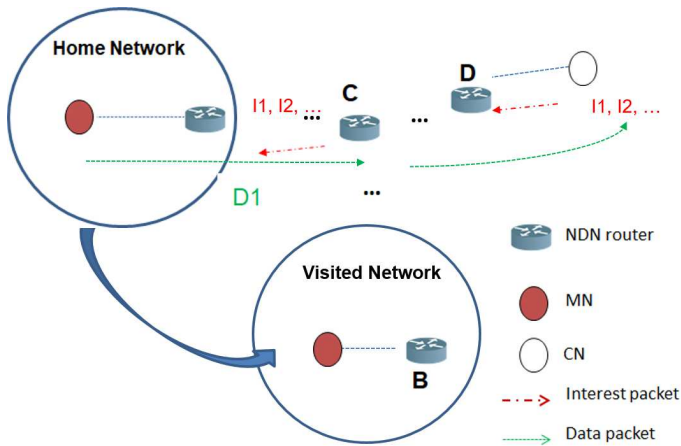}

\caption{Producer mobility example. The MN is a producer node on the move. Some data packets are sent before handing over. Once the MN attaches to the visited network, no data packets are sent, as the MN does not get Interest packets.\label{fig:producermobility}}
\end{figure}

Producer mobility is still a major challenge for ICN. To better
exemplify the issues with producer mobility, let us consider the scenario
previously addressed and illustrated in Figure \ref{fig:producermobility}, where $MN$, after receiving
an Interest packet $I1$ from its respective $CN$ forwarded to MN via router $A$, sends
back packet $D1$. MN then starts a handover to router $B$. In this
situation and again depending on the topology, the CN will keep on sending Interest packets to get subsequent data chunks. Routers in between shall keep on looking up their FIBs and as there is already an entry towards the respective
Name Prefix, routers shall forward the CN Interest packets towards the respective
Face (to router A). This process can result in significant latency. To circumvent this issue there is the need to rely on additional mobility management solutions.

Producer mobility is currently being handled via \emph{anchor-based approaches and anchor-less} approaches, as illustrated in Figure \ref{fig:producermobility}. Anchor-based
mechanisms, of which the most relevant is KITE~\cite{Afanasyev2016,Zhang2014}, follow IP-based approaches
and often recur to the use of a ``Rendez-Vous'' (RV) functional entity to temporarily assist data transmission. 
KITE  tries to exploit the forwarding states to keep track of nodes in movement. KITE considers that applications can send Interest packets to a routable and static anchor entity (an RV) to create the PIT entries as breadcrumbs. The RV is therefore a mediating entity, host-driven. Via this RV-based approach, Interest packets can reach a producer on the move.

This approach requires additional structures in routers -
a separate FIB or PIT - as well as additional state to be kept. Furthermore,
in current approaches the RV is considered to be static - mobility
of the RV is not handled. Therefore, while such approaches may be relevant from a perspective
of interoperability towards IP-based mobility solutions, the overhead
introduced can be significant. 

In what concerns anchor-less solutions, producers push a notification
once a move occurs. Such notification can be based on Interest packets
or on Data packets, being currently the preferred choice to rely on
a special Interest packet known as \emph{Interest Update}. This is
the case, for instance, of MAP-ME \cite{Auge2018}, or of MobiCCN
\cite{Wang2013}. \emph{Interest Updates} allow for arbitrary small
data to be placed in Interest packets as a name component. Such packet
is not registered in PITs, as no data is expected to be sent back.
Time-to-completion can be reduced by relying on different strategies,
such as occurs in MobiCCN, where a specific Name Prefix ``\emph{greedy:/}''
supports communication based on a greedy routing protocol. Or, a \emph{make-before-break}
approach can be followed, as occurs in the \emph{Publisher Mobility Support in Content-centric Networks (PMC)} solution~\cite{DookyoonHan2014}.

\begin{figure}
\centering
\includegraphics[scale=0.4]{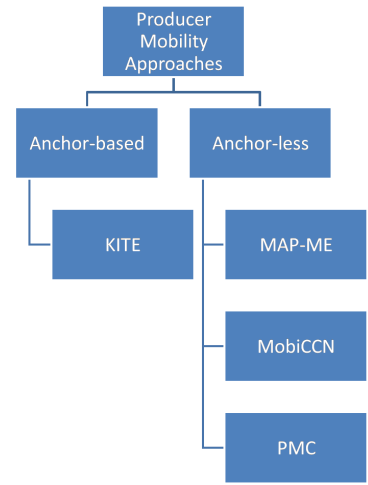}

\caption{Categorisation of mobility management approaches that provide support for producer mobility.}

\end{figure}

\section{Moving Towards Content-centric Mobility Management}
\label{sec:guidelines}

The benefits provided by the ICN architectural design in regards to
mobility support are the basis to rethink mobility management widely, having in mind a data-centric (and not host-centric) goal. 

ICN de-centralised, asynchronous and pull-based model removes the
need for a functional centralised or de-centralised MAP. Its architecture
can support consumer mobility naturally; however, there is still the
need to understand performance impact derived from the type of movement
as well as from the types of underlying topologies. 
In what concerns consumer mobility,
the pull-based nature of ICN gives the means to prevent serious packet
loss; nevertheless, latency impact is still not clear, and requires future work on performance aspects under highly variable scenarios.

Producer mobility, on the other hand, is still a challenge to be overcome. Related work
argues that producer mobility is a small subset of mobility. That
has been the case up until recently. With the advent of IoT and with
the growth of environments involving autonomous vehicles, producer
mobility becomes as relevant as consumer mobility. 

\subsection{The Relevancy of Context-aware Proactive Caching}
Proactive strategies for in-network caching can assist both consumer
and producer mobility, as  they support \emph{
make-before-break strategies} i.e., before a handover takes place.
While in-network caching approaches per se may not suffice to support mobility
\cite{Auge2018},proactive caching can be coupled with an anchor-less strategy to improve mobility support. The key aspect to consider in such approaches is to decide when
and where to cache content. Furthermore, reactive caching approaches are useful
in the context of host-oriented ICN mobility management approaches, as they assist in reducing packet loss while a node transits to a new location. It should be highlighted that while in IP-based solutions caching is used in regards to the first router in the path, in ICN caching refers to the content of the moving node and/or NDN routers in between.

Having in mind the support of data-oriented mobility management, proactive
approaches assist in caching the node's content before a handover occurs
(make-before-break). The data to cache and when to cache it can benefit
from mobility anticipation mechanisms as well as from network context
awareness, e.g., history of requests and producer neighbourhood context
\cite{Vasilakos2012,Lehmann2016}. 

Measures of neighbour availability and centrality, as well as measures
concerning similarity (for instance, similarity in types of requests),
and mobility awareness (e.g., handover frequency; estimation of time-to-handover)
can be easily provided via an external agent \cite{Sofia2017}. Such
information can assist in anticipating handovers, and selecting beforehand
a ``best'' neighbour to cache producer content. 

\subsection{Guidelines}

ICN is a relevant architecture to be integrated into large-scale mobile environments. While current mobility management proposals aim at solving specific issues under specific scenarios, for instance, producer mobility, future solutions need to consider the following:
\begin{itemize}
\item Producer and consumer mobility do not necessarily need to be treated independently, as has been done (by necessity) in prior approaches, which handled MN and CN mobility recurring to distinct mechanisms. In other words, the process of handling handovers should be the same for any node: any node becomes a MN. This can be supported by adding push-based communication support to ICN, via handover anticipation, for instance. 
\item Mobility anticipation mechanisms derived from context-awareness can be based on a MN's/CN's prior history and neighbourhood. Such concept is relevant to assist make-before-break handovers, thus eventually reducing the required signalling. In such cases producers 
can perform data push towards a ``best'' neighbour based on a proactive caching strategy. Via this mechanism, packet loss can be reduced at the expense of a (potentially) small increase in overhead. 
\item The relevant aspect in an ICN context is ``when'' a move may occur, and not ``where to'' the node shall move. ICN provides global naming, so the location where nodes are should not be the key aspect, from a content-centric mobility management perspective. 
\item A proactive caching approach towards a ``best'' neighbour of a node can benefit from being associated with a specific Name Prefix, or specific metadata associated with the content to be transmitted. For instance, a timeout (TTL, TLV), or priority numbering.
\item Once a move occurs, nodes should emit a notification. While this is the common procedure for consumers, producers can emit an Interest Update notification as envisioned in the original ICN/CCNx design. This notification allows for a faster routing re-establishment.
\item Naming in ICN is hierarchical and independent of location. Nevertheless,
today it is common to consider a naming space associated with routing
domains, e.g. ``/lusofona.pt/videos/''. While such choice does not
impact mobility management, it may negatively impact route aggregation, when
producers move. ICN applications would benefit from a set of guidelines
for the development of the naming space.
\end{itemize}

\section{Conclusions and Future Research Directions}
\label{sec:conclusions}

The benefits provided by the intrinsic ICN architectural design in regards to mobility are
the basis to rethink mobility management widely and from a content-centric
perspective. The ICN architectural design removes the need for a functional
centralised or de-centralised mobility anchor-point. As such, anchor-less
approaches seem to be a relevant approach as they reduce the need
for additional state, and allow ICN to support mobility management
in a data-centric way. 

Consumer mobility is well supported from a network architectural perspective,
but there is the need to understand performance impact derived from
the type of movement as well as from the types of topologies. The
pull-based nature of ICN gives the means to prevent serious packet
loss; nevertheless, consumer mobility may still result in large time-to-completion
intervals.

In what concerns producer mobility, the support is not intrinsic,
and the receiver-driven, pull-based ICN approach requires adjustments
to fully support producer mobility. Multi-homing is well support.

While ICN has relevant architectural properties which seem to provide a better and integrated support for mobility management, there are a few aspects requiring further research. 
A first future research direction concerns a better support for mobility management via NDN routing. By devising routing approaches that are sensitive to node movement~ \cite{Chama2015}, Interest packets can be forwarded in a way that is automatically based on individual and collective roaming habits of devices, eventually reducing the need to perform re-registration once nodes reattach. This can be done by developing forwarding strategies based on mobility prediction, or by integrating routing support based on context-awareness~\cite{mendes-icnrg-dabber-02}.
A second research direction is to perform an analysis of the ICN mobility support support in highly variable topologies, where anchor-less strategies may not suffice to adequately support producer mobility. For this case, both producer and consumer need to be considered as mobile entities and therefore, any future mobility management approach should simply look into the support of a single mobile entity, instead of supporting, as previously, a MN and a CN entity separation.

\vspace{6pt} 




\section*{Acknowledgments}
The author thanks the insightful exchange of ideas concerning ICN and IoT held in 2017/2018 with Chris Winkler (Siemens AG); Hans-Peter Huth (Siemens AG); Jan Seeger (Siemens AG and Technical University of Munich); Georg Carle (Technical University
of Munich).




\bibliographystyle{IEEEtran}

\bibliography{MobilityICN.bib,Mobility2019.bib}

\begin{thebibliography}{10}
\providecommand{\url}[1]{#1}
\csname url@samestyle\endcsname
\providecommand{\newblock}{\relax}
\providecommand{\bibinfo}[2]{#2}
\providecommand{\BIBentrySTDinterwordspacing}{\spaceskip=0pt\relax}
\providecommand{\BIBentryALTinterwordstretchfactor}{4}
\providecommand{\BIBentryALTinterwordspacing}{\spaceskip=\fontdimen2\font plus
\BIBentryALTinterwordstretchfactor\fontdimen3\font minus
  \fontdimen4\font\relax}
\providecommand{\BIBforeignlanguage}[2]{{%
\expandafter\ifx\csname l@#1\endcsname\relax
\typeout{** WARNING: IEEEtran.bst: No hyphenation pattern has been}%
\typeout{** loaded for the language `#1'. Using the pattern for}%
\typeout{** the default language instead.}%
\else
\language=\csname l@#1\endcsname
\fi
#2}}
\providecommand{\BIBdecl}{\relax}
\BIBdecl

\bibitem{Ravindran}
\BIBentryALTinterwordspacing
R.~Ravindran, A.~Chakraborti, S.~O. Amin, and A.~Azgin, ``{5G-ICN : Delivering
  ICN Services over 5G using Network Slicing},'' \emph{IEEE Communications
  Magazine 55.5}, pp. 101--107, 2017. [Online]. Available:
  \url{https://arxiv.org/pdf/1610.01182.pdf}
\BIBentrySTDinterwordspacing

\bibitem{Afanasyev2016}
\BIBentryALTinterwordspacing
A.~Afanasyev, J.~Burke, and L.~Zhang, ``{A Survey of Mobility Support in Named
  Data Networking},'' \emph{inProc. IEEE Conference on Computer Communications
  Workshops (INFOCOM WKSHPS)}, pp. 83--88, 2016. [Online]. Available:
  \url{http://irl.cs.ucla.edu/data/files/papers/nom16-ndn-mobility-survey.pdf}
\BIBentrySTDinterwordspacing

\bibitem{Chen2016mobility}
S.~Chen, Y.~Shi, B.~Hu, and M.~Ai, \emph{Mobility Management: Principle,
  Technology and Applications}.\hskip 1em plus 0.5em minus 0.4em\relax
  Springer, 2016.

\bibitem{dmm}
\BIBentryALTinterwordspacing
D.~Liu, J.~Zuniga, P.~Seite, H.~Chan, and C.~Bernardos, ``{Distributed Mobility
  Management: Current Practices and Gap Analysis},'' \emph{IETF DMM Working
  Group, RFC 7429}, 2015. [Online]. Available:
  \url{http://tools.ietf.org/wg/dmm/}
\BIBentrySTDinterwordspacing

\bibitem{Liu2014}
\BIBentryALTinterwordspacing
D.~Liu, P.~Seite, H.~Yokota, and J.~Korhonen, ``{Requirements for Distributed
  Mobility Management},'' \emph{IETF DMM Working Group, RFC 7429}, aug 2014.
  [Online]. Available: \url{https://www.rfc-editor.org/info/rfc7333}
\BIBentrySTDinterwordspacing

\bibitem{Zhu2013a}
\BIBentryALTinterwordspacing
Z.~Zhu, A.~Afanasyev, and L.~Zhang, ``{A new perspective on mobility
  support},'' \emph{Named Data Networking Project Technical Report 13, USA},
  pp. 1--6, 2013. [Online]. Available:
  \url{http://new.named-data.net/wp-content/uploads/TRmobility.pdf}
\BIBentrySTDinterwordspacing

\bibitem{Auge2018}
\BIBentryALTinterwordspacing
J.~Auge, G.~Carofiglio, G.~Grassi, L.~Muscariello, G.~Pau, and X.~Zeng,
  ``{MAP-Me: Managing Anchor-Less Producer Mobility in Content-Centric
  Networks},'' \emph{IEEE Transactions on Network and Service Management},
  vol.~15, no.~2, pp. 596--610, jun 2018. [Online]. Available:
  \url{https://ieeexplore.ieee.org/document/8267132/}
\BIBentrySTDinterwordspacing

\bibitem{Zhang2014}
\BIBentryALTinterwordspacing
Y.~Zhang, H.~Zhang, and L.~Zhang, ``{Kite: A mobility support scheme for
  NDN},'' \emph{inProc. of the 1st international conference on
  Information-centric networking - INC '14}, pp. 179--180, 2014. [Online].
  Available: \url{http://dl.acm.org/citation.cfm?doid=2660129.2660159}
\BIBentrySTDinterwordspacing

\bibitem{Chen2014}
S.~Chen, Y.~Shi, B.~Hu, and M.~Ai, ``{Mobility-driven Networks (MDN): From
  Evolution to Visions of Mobility Management},'' \emph{IEEE Network}, vol.~28,
  no.~4, pp. 66--73, 2014.

\bibitem{tyson2012survey}
G.~Tyson, N.~Sastry, I.~Rimac, R.~Cuevas, and A.~Mauthe, ``A survey of mobility
  in information-centric networks: Challenges and research directions,'' in
  \emph{Proceedings of the 1st ACM workshop on Emerging Name-Oriented Mobile
  Networking Design-Architecture, Algorithms, and Applications}.\hskip 1em plus
  0.5em minus 0.4em\relax ACM, 2012, pp. 1--6.

\bibitem{Nascimento2011}
A.~Nascimento, R.~Sofia, T.~Condeixa, and S.~Sargento, ``{A Characterization of
  Mobility Management in User-centric Networks},'' in \emph{inProc. of the 11th
  international conference and 4th international conference on Smart spaces and
  next generation wired/wireless networking}, ser. NEW2AN'11/ruSMART'11.\hskip
  1em plus 0.5em minus 0.4em\relax Berlin, Heidelberg: Springer-Verlag, 2011,
  pp. 314--325.

\bibitem{rfc4885}
T.~Ernst and H.~Lach, ``Network mobility support terminology,'' \emph{IETF
  Network Working Group, RFC 4885 (informational).}, 2007.

\bibitem{Wong2007}
\BIBentryALTinterwordspacing
K.~D. Wong, A.~Dutta, H.~Schulzrinne, and K.~Young, ``{Simultaneous mobility:
  Analytical framework, theorems and solutions},'' \emph{Wireless
  Communications and Mobile Computing}, vol.~7, no.~5, pp. 623--642, 2007.
  [Online]. Available:
  \url{http://www.cs.columbia.edu/{~}dutta/research/simmob-wcmc.pdf}
\BIBentrySTDinterwordspacing

\bibitem{Schneider2015}
\BIBentryALTinterwordspacing
K.~M. Schneider, K.~Mast, and U.~R. Krieger, ``{CCN forwarding strategies for
  multihomed mobile terminals},'' in \emph{inProc. International Conference and
  Workshops on Networked Systems (NetSys)}.\hskip 1em plus 0.5em minus
  0.4em\relax IEEE, mar 2015, pp. 1--5. [Online]. Available:
  \url{http://ieeexplore.ieee.org/document/7089075/}
\BIBentrySTDinterwordspacing

\bibitem{mipv6}
\BIBentryALTinterwordspacing
C.~E. Perkins, D.~B. Johnson, and J.~Arkko, ``{IETF RFC 6275 - Mobility Support
  in IPv6},'' \emph{IETF Network Working Group RFC 6275}, 2011. [Online].
  Available: \url{https://tools.ietf.org/html/rfc6275}
\BIBentrySTDinterwordspacing

\bibitem{sip}
\BIBentryALTinterwordspacing
J.~Rosenberg, H.~Schulzrinne, G.~Camarillo, A.~Johnston, J.~Peterson,
  R.~Sparks, M.~Handley, and E.~Schooler, ``{Session Initiation Protocol},''
  \emph{IETF Network Working Group RFC 3261}, jun 2002. [Online]. Available:
  \url{http://www.ietf.org/rfc/rfc3261.txt}
\BIBentrySTDinterwordspacing

\bibitem{Nikander}
\BIBentryALTinterwordspacing
P.~Nikander and R.~Moskowitz, ``{ Host Identity Protocol (HIP) Architecture},''
  \emph{{ IETF Network Working Group RFC 4423}}, 2006. [Online]. Available:
  \url{https://tools.ietf.org/html/rfc4423}
\BIBentrySTDinterwordspacing

\bibitem{koh2005}
\BIBentryALTinterwordspacing
S.~J. Koh, ``{Mobile SCTP for IP Mobility Support in All-IP Networks},'' in
  \emph{Proceedings of CIC (Cellular and Intelligent Communications)}, 2003.
  [Online]. Available:
  \url{https://www.semanticscholar.org/paper/Mobile-SCTP-for-IP-Mobility-Support-in-All-IP-Koh/7372180d4fe83b18bd83425bcd28a05ae2bcf28d}
\BIBentrySTDinterwordspacing

\bibitem{koponen2007}
\BIBentryALTinterwordspacing
T.~Koponen, M.~Chawla, B.-G. Chun, A.~Ermolinskiy, K.~H. Kim, S.~Shenker, and
  I.~Stoica, ``{A data-oriented (and beyond) network architecture},'' \emph{ACM
  SIGCOMM Computer Communication Review}, vol.~37, no.~4, p. 181, 2007.
  [Online]. Available:
  \url{http://portal.acm.org/citation.cfm?doid=1282427.1282402}
\BIBentrySTDinterwordspacing

\bibitem{Gusev2017}
\BIBentryALTinterwordspacing
P.~Gusev and J.~Burke, ``{CICN - Content Centric networking Community},'' 2017.
  [Online]. Available:
  \url{http://dl.acm.org/citation.cfm?doid=2810156.2810176}
\BIBentrySTDinterwordspacing

\bibitem{Dannewitz2013}
\BIBentryALTinterwordspacing
C.~Dannewitz, D.~Kutscher, B.~Ohlman, S.~Farrell, B.~Ahlgren, and H.~Karl,
  ``{Network of information (NetInf)-An information-centric networking
  architecture},'' \emph{Computer Communications}, vol.~36, no.~7, pp.
  721--735, 2013. [Online]. Available:
  \url{http://dx.doi.org/10.1016/j.comcom.2013.01.009
  https://www.it.uu.se/edu/course/homepage/projektDV/ht14/NetInfArch-elsevier-published.pdf}
\BIBentrySTDinterwordspacing

\bibitem{dimitrov2010}
\BIBentryALTinterwordspacing
V.~Dimitrov and V.~Koptchev, ``{PSIRP project -- publish-subscribe internet
  routing paradigm},'' in \emph{Proceedings of the 11th International
  Conference on Computer Systems and Technologies and Workshop for PhD Students
  in Computing on International Conference on Computer Systems and Technologies
  - CompSysTech '10}.\hskip 1em plus 0.5em minus 0.4em\relax New York, New
  York, USA: ACM Press, 2010, p. 167. [Online]. Available:
  \url{http://portal.acm.org/citation.cfm?doid=1839379.1839409}
\BIBentrySTDinterwordspacing

\bibitem{Tyson2012}
\BIBentryALTinterwordspacing
G.~Tyson, A.~Mauthe, S.~Kaune, P.~Grace, A.~Taweel, and T.~Plagemann,
  ``{Juno},'' \emph{ACM Transactions on Internet Technology}, vol.~12, no.~2,
  pp. 1--28, 2012. [Online]. Available:
  \url{http://dl.acm.org/citation.cfm?doid=2390209.2390210}
\BIBentrySTDinterwordspacing

\bibitem{zhang2010}
\BIBentryALTinterwordspacing
L.~Zhang, D.~Estrin, J.~Burke, V.~Jacobson, J.~D. Thornton, D.~K. Smetters,
  B.~Zhang, G.~Tsudik, D.~Massey, C.~Papadopoulos, L.~Wang, P.~Crowley, and
  E.~Yeh, ``{Named Data Networking (NDN) Project},'' \emph{October}, pp. 1--26,
  2010. [Online]. Available:
  \url{http://www.cl.cam.ac.uk/{~}ey204/teaching/ACS/R202/papers/S3{\_}CCN{\_}NDN/papers/zhang{\_}ndn{\_}2010.pdf}
\BIBentrySTDinterwordspacing

\bibitem{Wang2013}
\BIBentryALTinterwordspacing
L.~Wang, O.~Waltari, and J.~Kangasharju, ``{MobiCCN: Mobility support with
  greedy routing in Content-Centric Networks},'' \emph{inProc. GLOBECOM - IEEE
  Global Telecommunications Conference}, pp. 2069--2075, 2013. [Online].
  Available: \url{https://www.cs.helsinki.fi/u/lxwang/publications/mobiccn.pdf}
\BIBentrySTDinterwordspacing

\bibitem{DookyoonHan2014}
\BIBentryALTinterwordspacing
{Dookyoon Han}, M.~Lee, K.~Cho, T.~T. Kwon, and Y.~Cho, ``{Publisher Mobility
  Support in Content Centric Networks},'' \emph{inProc. Information Networking
  (ICOIN), 2014 International Conference on. IEEE, 2014}, pp. 214--219, 2014.
  [Online]. Available:
  \url{https://www.researchgate.net/publication/269310003{\_}Publisher{\_}mobility{\_}support{\_}in{\_}content{\_}centric{\_}networks}
\BIBentrySTDinterwordspacing

\bibitem{Vasilakos2012}
\BIBentryALTinterwordspacing
X.~Vasilakos, V.~A. Siris, G.~C. Polyzos, and M.~Pomonis, ``{Proactive
  selective neighbor caching for enhancing mobility support in
  information-centric networks},'' \emph{Proceedings of the second edition of
  the ICN workshop on Information-centric networking - ICN '12}, p.~61, 2012.
  [Online]. Available:
  \url{http://dl.acm.org/citation.cfm?doid=2342488.2342502}
\BIBentrySTDinterwordspacing

\bibitem{Lehmann2016}
\BIBentryALTinterwordspacing
M.~B. Lehmann, M.~P. Barcellos, and A.~Mauthe, ``{Providing producer mobility
  support in NDN through proactive data replication},'' \emph{inProc. of the
  NOMS 2016 - 2016 IEEE/IFIP Network Operations and Management Symposium}, pp.
  383--391, 2016. [Online]. Available:
  \url{https://core.ac.uk/download/pdf/42415808.pdf}
\BIBentrySTDinterwordspacing

\bibitem{Sofia2017}
\BIBentryALTinterwordspacing
R.~C. Sofia, I.~Santos, J.~Soares, S.~Diamantopoulos, C.-A. Sarros,
  D.~Vardalis, V.~Tsaoussidis, and A.~D'Angelo, ``{UMOBILE D4.5: Report on Data
  Collection and Inference Models},'' COPELABS, University Lusofona; Senception
  Lda, Tech. Rep., sep 2017. [Online]. Available:
  \url{http://siti2.ulusofona.pt:8085/xmlui/handle/20.500.11933/733}
\BIBentrySTDinterwordspacing

\bibitem{Chama2015}
N.~Chama, R.~C. Sofia, and S.~Sargento, ``{A Discussion on Mobility Awareness
  of Multi Hop Routing in User-Centric Environments},'' COPELABS, University
  Lus{\{}{\'{o}}{\}}fona, University of Aveiro, Tech. Rep. COPE-SITI-TR-05-15,
  2015.

\bibitem{mendes-icnrg-dabber-02}
\BIBentryALTinterwordspacing
P.~Mendes, R.~C. Sofia, V.~Tsaoussidis, S.~Diamantopoulos, C.-A. Sarros,
  C.~Borrego, and J.~Borrell, ``{Information-centric Routing for Opportunistic
  Wireless Networks},'' Internet Engineering Task Force, Internet-Draft
  draft-mendes-icnrg-dabber-02, 2019, work in Progress. [Online]. Available:
  \url{https://datatracker.ietf.org/doc/html/draft-mendes-icnrg-dabber-02}
\BIBentrySTDinterwordspacing

\end{thebibliography}

\end{document}